\documentclass[aps,prl,twocolumn,superscriptaddress,showpacs]{revtex4-1}
\usepackage{graphicx}
\usepackage[german,english]{babel}
\usepackage{amssymb}
\usepackage{amsmath}
\usepackage{braket}
\usepackage{subfigure}

\newcommand{\eff}{\ensuremath{{\rm eff}}}
\newcommand{\up}{\ensuremath{\uparrow}}
\newcommand{\dn}{\ensuremath{\downarrow}}

\begin{document}

\title{Excitation spectra of transition metal atoms on the Ag (100) surface controlled by Hund's exchange}

\author{S. Gardonio}
\affiliation{University of Nova Gorica Vipavska 11c, 5270 Ajdov\v{s}\v{c}ina, Slovenia}

\author{M. Karolak}
\email{mkarolak@physik.uni-hamburg.de}
\affiliation{I. Institut f{\"u}r Theoretische Physik, Universit{\"a}t Hamburg, Jungiusstra{\ss}e 9, D-20355 Hamburg, Germany} 

\author{T. O. Wehling}
\email{Corresponding author.\\ wehling@itp.uni-bremen.de}
\affiliation{Institut f{\"u}r Theoretische Physik, Universit{\"a}t Bremen, Otto-Hahn-Allee 1, D-28359 Bremen, Germany }
\affiliation{Bremen Center for Computational Materials Science, Universit{\"a}t Bremen, Am Fallturm 1a, 28359 Bremen, Germany}

\author{L. Petaccia} 
\author{S. Lizzit}
\author{A. Goldoni} 
\affiliation{Elettra Sincrotrone Trieste, Strada Statale 14 km 163.5, 34149 Trieste, Italy}

\author{A. I. Lichtenstein}
\affiliation{I. Institut f{\"u}r Theoretische Physik, Universit{\"a}t Hamburg, Jungiusstra{\ss}e 9, D-20355 Hamburg, Germany}

\author{C. Carbone}
\affiliation{Istituto di Struttura della Materia, Consiglio Nazionale delle Ricerche, Trieste, Italy}


\date{\today}

\begin{abstract}
We report photoemission experiments revealing the valence electron spectral function of Mn, Fe, Co and Ni atoms on the Ag(100) surface. The series of spectra shows splittings of higher energy features which decrease with the filling of the $3d$ shell and a highly non-monotonous evolution of spectral weight near the Fermi edge. First principles calculations demonstrate that two manifestations of Hund's exchange $J$ are responsible for this evolution. First, there is a monotonous reduction of the effective exchange splittings with increasing filling of the $3d$ shell. Second, the amount of charge fluctuations and, thus, the weight of quasiparticle peaks at the Fermi level varies non-monotonously through this $3d$ series due to a distinct occupancy dependence of effective charging energies $U_\eff$.

\end{abstract}

\maketitle

Strongly correlated multi-orbital quantum systems present a classical yet unsolved problem in solid state physics which appears in various systems ranging from periodic solids to isolated atoms on surfaces.
In nanoscopic systems, atomic scale control of correlated electrons holds promises for novel modes of information processing \cite{Khajetoorians_Science2011,Loth_Science2012} and offers possibilities to understand fundamental quantum effects such as itinerant electron magnetism, the competition of local with non-local magnetic interactions \cite{Kroeger_PRL11,Wahl_2Kondo2011,HewsonBook} or the transition from isolated atoms to extended solids \cite{Gambardella_review2009,carbone_2010,Ce_PRL2011}. Here, the multi orbital nature of realistic nanoscale transition metal (TM) structures is generally believed to control physical properties such as magnetic anisotropies \cite{Gambardella_review2009}, magnetic excitations \cite{Otte_Kondo_Anisotropy2008} or Kondo temperatures \cite{nevidomskyy_2009}. 

While there has been huge the experimental progress in fabricating nanoscale correlated electron systems, our theoretical understanding of the physical mechanisms determining their electronic and magnetic properties is still at a rather basic qualitative level: Often experiments on transition metal impurity systems are interpreted in terms of Kondo type models where a generalized spin is coupled to a bath of conduction electrons \cite{HewsonBook,costi_1996,Otte_Kondo_Anisotropy2008} or in terms of single orbital Anderson models \cite{Zawadowski_PRL00,Kern02,Wenderoth04,BJones_PRL06,Neel_08,Ternes_Heinrich_Schneider_Review09,Wahl_NJP09,Wenderoth11}. However, links between these models and \textit{realistic} systems are usually ambiguous, difficult to establish \cite{Costi_09} and can typically only be made a posteriori \cite{Zawadowski_PRL00,Kern02,Wenderoth04,BJones_PRL06,Neel_08,Ternes_Heinrich_Schneider_Review09,Wahl_NJP09,Wenderoth11}.
It is often largely unclear which microscopic degrees of freedom are active at a given energy scale and how their 
contribution in excitation spectra measured with different spectroscopy techniques can be disentangled. 

In this letter, we consider the series of isolated Mn, Fe, Co and Ni adatoms on the Ag (100) surface and explain their excitation spectra. We report on valence electron photoemission experiments revealing a complex evolution of the electronic spectra through this series: we find a monotonous decrease in the splitting of higher energy features and a non-monotonous variation of low energy spectral weight. By means of first-principles calculations we explain the photoemission results and show that both observations trace back to Hund's exchange. First, the splitting between final state multiplets with different spin decreases monotonously due to a monotonous reduction of effective exchange splittings $J(n_\up-n_\dn)$ with increasing filling of the $3d$ shell. Second, the effective charging energies \cite{Werner_2009,de_medici_PRB2011,de_medici_2011} $U_\eff(n)=E(n+1)+E(n-1)-2E(n)$, with $E(n)$ being the ground state energy of the impurity adatom with $n$ electrons, vary due to Hund's exchange in a strongly non-monotonous way from Mn to Ni. Therefore, the amount of charge fluctuations and the weight of quasiparticle peaks at the Fermi level evolves non-monotonously through this $3d$ series. We find sizable charge fluctuations and mixed valence behavior for Fe and Ni. In contrast Mn and Co come closer to a multi-orbital Kondo limit with a generalized impurity spin being coupled to a bath of conduction electrons and less charge fluctuations.

\begin{figure}
\centering
\includegraphics[width=\linewidth]{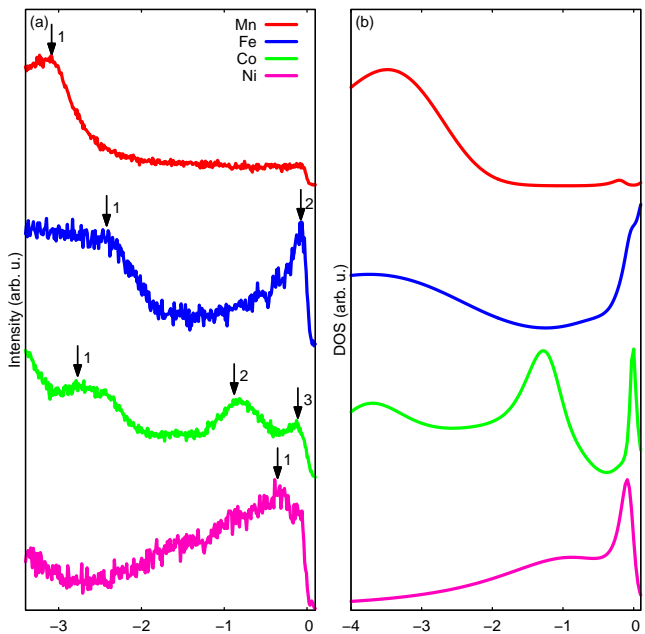}
\caption{(Color online) Valence band spectra of Mn, Fe, Co, and Ni adatoms (from top to bottom) of Ag (100). (a) Experimental photoemission spectra (b) Theoretical spectra obtained from QMC calculations at $\beta=20$eV$^{-1}$ with imaginary time discretization of $\Delta\tau=0.125$ via analytical continuation \cite{maxent}. The $3d$ shell occupancies used in the simulations are $n=5.0$ for Mn, $n=6.0$ for Fe, $n=7.8$ for Co, and $n=8.4$ for Ni. }
\label{fig_1}
\end{figure}

For our experiments, we prepared the Ag(100) substrate by the standard procedure and monitored its crystalline quality by low energy electron diffraction (LEED). The LEED pattern quality was very high, with sharp diffraction spots on low background. Isolated TM adatoms were obtained by depositing Mn, Fe, Co and Ni atoms at a substrate temperature of 20 K (statistical growth regime). The TM coverages were calibrated by a quartz microbalance and here we conventionally define the coverage according to mass equivalent of a nominal monolayer. The sample temperature was maintained at 20 K during the photoemission measurements.
The photoemission spectra were measured with a photon energy of 120eV which corresponds to the Cooper minimum of the Ag $4d$ photoionization cross section. Under these experimental conditions, there is an enhancement of the signal of the impurity (coverages in the range of $2\times10^{-2}$ to $10^{-3}$ monolayers) with respect to the one of the host surface. The photoemission experiments were performed at the SuperESCA beamline at the ELETTRA Synchrotron Radiation facility, with an overall energy resolution better than 40 meV. 

To explain the experimental spectra we performed density functional theory (DFT) calculations using the generalized gradient approximation (GGA) \cite{Perdew:PW91} as implemented in the Vienna Ab-Initio Simulation Package (VASP) \cite{Kresse:PP_VASP} with the projector augmented waves basis sets (PAW) \cite{Kresse:PAW_VASP,Bloechl:PAW1994}. In these calculations, single transition metal atoms on Ag(100) have been modeled using $4\times 4$ surface supercells with slab thicknesses of five layers. The crystal structures have been relaxed until the forces acting on each atom were less than 0.01eV/\AA. This yields hybridization functions of the adatoms, which are then used to define five orbital Anderson impurity models with Coulomb interactions given through the Slater integrals $F^0=U$, $F^2=14/(1 + 0.625)J$ and $F^4= 0.625F^2$ \cite{Anisimov_1997} with $U=3$\,eV for Mn and Fe ($U=5$\,eV for Co and Ni) as well as Hund's exchange $J=1$\,eV. As the occupancies $n$ of the $3d$ impurity orbitals are not exactly known, they are kept as free parameters. The impurity models were solved using the Hirsch-Fye Quantum Monte Carlo method (QMC) \cite{hf_qmc} (keeping the density-density part of the local Coulomb interaction) as well as exact diagonalization (ED) \cite{hubbard_one}. In this way, we obtain the adatom spectra including electron correlation effects in QMC as well as a detailed insight into the atomic multiplet structure via ED.

Fig. \ref{fig_1} (a) shows the experimental photoelectron energy distribution curves in the valence band of isolated Mn, Fe, Co and Ni atoms on the Ag(100) surface. The curves are difference spectra between the clean Ag surface and the surface covered with a few adatoms and thus correspond to the contribution from $3d$ impurity electronic states \cite{supplemental}. We observe a remarkable evolution of the impurity spectra through this series of TM adatoms: Mn possess one structure (labeled 1) at binding energy (BE) of 3.25 eV; Fe has two structures, one (1) at 2.32 eV BE  and the other one (2) near the Fermi Energy ($E_F$); Co has one (1) broad structure at 2.57 eV BE, one structure (2) at 0.8 eV BE, and one structure (3) close to the $E_F$; and Ni has one broad structure (1) at 0.35eV BE.

We start with reconciling these experimental results in the context of a generalized Kondo description: For Mn, the spectral peak at $-3.25$\,eV and virtually no quasiparticle peak at the Fermi level could be well in line with Mn acting effectively as a spin $S=5/2$ Kondo impurity. Indeed, this would be very similar to the situation found for Mn impurities in bulk Ag, which has been derived from photoemission spectroscopy and measurements of the magnetic susceptibility \cite{HewsonBook}. The virtually absent quasiparticle peak would then be well understandable as the large spin $S=5/2$ leads to very low Kondo temperatures \cite{nevidomskyy_2009}. With increasing filling of the $3d$ shell the impurity spin should be gradually reduced and the spectral weight of the quasiparticle peak near the Fermi level should be growing exponentially. Indeed, Fe, Co, and Ni exhibit spectral weight near the Fermi level but the shape and weight of these low energy spectral varies very non-monotonically through the series of Fe, Co and Ni. In particular, we do not find a monotonous increase of the quasiparticle spectral weight as would be expected in spin-only Kondo models \cite{nevidomskyy_2009}. Thus, we conclude that additional degrees of freedom must be responsible for the experimentally observed evolution of the adatom spectra. To pinpoint these degrees of freedom and to explain the spectra theoretically, we present calculations combining density functional theory (DFT) and quantum many body methods in the following.

The DFT calculations show that all transition metal adatoms adsorb to high symmetry positions continuing the Ag lattice, i.e. sitting in the center of a square of Ag atoms. The adsorption height above the surface differs only little, from Mn at 1.30~\r{A} to Ni at approximately 1.4~\AA. In line with the similar adsorption geometries the hybridization functions, Im~$\Delta(\omega)$, of the adatoms are similar for all adatoms \cite{supplemental}. Most importantly, the hybridization function is rather featureless for all adatoms in the energy region between $-3$\,eV and $+1$\,eV. Thus, the complex evolution of the spectra observed experimentally also cannot be a single particle hybridization effect.

\begin{figure}
\centering
\includegraphics[width=\linewidth]{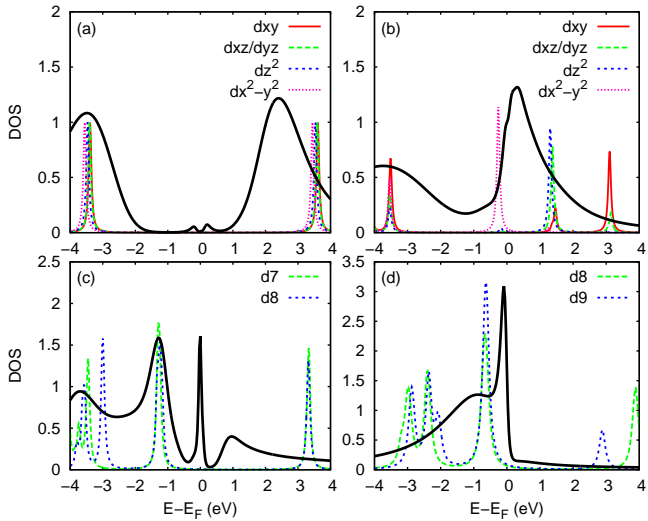}
\caption{(Color online) Spectral functions for (a) Mn, (b) Fe, (c) Co and (d) Ni impurities obtained from QMC calculations at $\beta=20$eV$^{-1}$ with imaginary time resolution of $\Delta\tau=0.125$ via analytical continuation \cite{maxent} (thick black lines). The $3d$ shell occupancies used in the simulations are $n=5$ for Mn, $n=6$ for Fe, $n=7.8$ for Co, and $n=8.4$ for Ni. Additionally spectra obtained by ED are shown (orbitally resolved in a) and b); total spectra for different filling of the $3d$ shell in c) and d)).}
\label{fig:spectra}
\end{figure}

The spectral functions of Anderson impurity models obtained from QMC and ED are shown in Fig. \ref{fig_1} (b) and in Fig. \ref{fig:spectra}. In agreement with our experiments, the Mn spectrum consists mainly of one peak far below the Fermi level for $3d$ shell fillings $n\approx 5$. In fact, already a diagonalization of the atom in the crystal field ($\Delta_{cf}$) of the surface shows the basic structure found in the experiment (Fig. \ref{fig:spectra}a). Thus, a low energy description of Mn on Ag (100) in terms of a spin $S=5/2$ Kondo model is well in line with our results. This is indeed similar to the case Mn in bulk Ag \cite{HewsonBook} and also in agreement with DFT calculations for Mn on Ag(100) \cite{nonas_2001,cabria_2002}. Comparison of the QMC calculations to the experimental spectra reveals good agreement also for the Fe, Co, and Ni adatoms. We thus use the QMC results to understand the physical mechanisms behind the evolution of the spectra in the series of $3d$ adatoms.

\begin{figure}
\centering
\includegraphics[width=\linewidth]{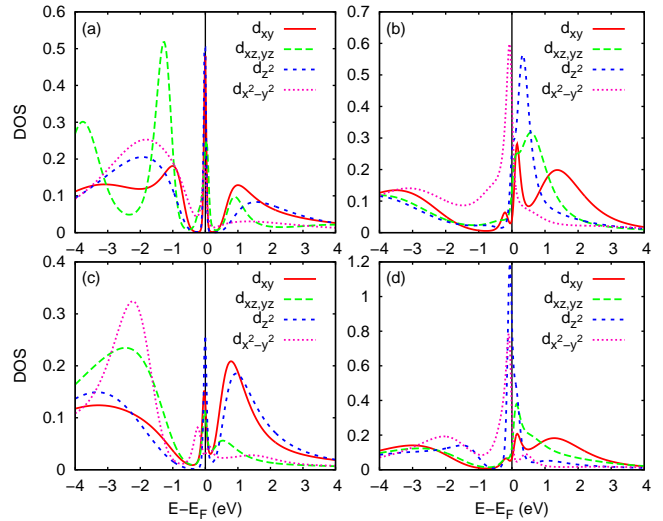}
\caption{\label{fig:orbdos}(Color online) Orbitally resolved spectral functions of Co ($n=7.8$ (a) and $n=7.9$ (c)) and Fe ($n=6.0$ (b) and $n=6.4$ (d)) impurities as obtained from QMC.}
\end{figure}

\textit{Multiplet splittings and Hund's exchange.} 
For Fe with $n=6$, our calculations show a broad peak around $-3$\,eV and a relatively narrow peak right below the Fermi level, which reproduce the experimental features 1 and 2, respectively. The broad satellite around $-3$\,eV appears in all orbitals and is found also in the ED calculations (Fig. \ref{fig:spectra}b). It can be identified as $d^6 \to d^5$ ionization peak. Analyzing the orbitally resolved spectral function shown in Fig. \ref{fig:orbdos}(b) we see that the experimental feature 2 stems from the $d_{x^2-y^2}$ orbital. This orbital has the occupancy 0.8 in contrast to the other orbitals which are approximately half filled. Therefore, the feature 2 cannot be a quasiparticle peak due to spin-only Kondo physics. 
Indeed, this peak appears already in an ED description of Fe in the crystal field (Fig. \ref{fig:spectra} (b)) of the surface, which shows that it corresponds to a $d^6 \to d^5$ transition. The spectral features 1 and 2 of Fe, thus, stem both from ionization processes of the impurity. Our ED calculations further show that the energy separation of these peaks traces back to different $d^5$ final state multiplets: $S=5/2$, $L=0$ for feature 2 near $E_F$ and higher energy multiplets such as $S=3/2$, $L\geq 0$ for feature 1. The splitting between these multiplets can be understood as an effective exchange splitting $\sim J(n_\up-n_\dn)$ due to Hund's exchange.

The experimental Co spectrum consisting of three peaks is well reproduced in our QMC simulations for $n=7.8$ (Fig.~\ref{fig:spectra}c). Analyzing the orbitally resolved spectral function shown in Fig.~\ref{fig:orbdos} (a,c) and the corresponding occupation matrices we find that the $d_{xz,yz}$ and $d_{x^2-y^2}$ orbitals are almost fully occupied and mostly responsible for the peak at $-1$\,eV, which corresponds to the feature 2 in the experimental spectra.
Again, the spectral weight further below the Fermi level (feature 1) traces also back to an ionization process, here $d^8 \to d^7$, with higher energy final state multiplets. As in the case of Fe, the splitting between the features 1 and 2 can thus be understood as atomic multiplet effect due to Hund's exchange. For Co, the effective exchange splitting reduces as compared to the case of Fe by an amount on the order of $J$. This explains why the separation of the peaks 1 and 2 is 0.5 eV smaller for Co than for Fe.

The spectrum of Ni turns out to consist mainly of a broad peak below the Fermi level without clearly resolvable multiplet features. This is qualitatively in line with an even further reduced effective exchange splitting $\sim J(n_\up-n_\dn)$ for Ni. Indeed, in our GGA and GGA+U calculations, the Ni adatoms turn out to be nonmagnetic on this surface which is in agreement with calculations in Ref. \cite{lazarovits_2002} and experiments for Ni on Au \cite{beckmann_1996}.

\textit{Effective charging energies and valence fluctuations.}
It remains to be explained why feature 2 in the Fe spectrum and the whole Ni spectral peak are rather close to $E_F$ and what the nature of the peak 3 near $E_F$ in the Co spectrum is. Therefore, we discuss the issues of valence fluctuations as well as Kondo physics in the following.

For a fully rotationally invariant Coulomb vertex there is a pronounced occupancy dependence of the effective charging energies \cite{supplemental,Haverkort_phd}
\begin{equation}
U_\eff(n)\approx\left\{ \begin{array}{crl}
U+4J & \text{for} & n=5\\
U-(3/2)J & \text{for} &n=6\; \text{and}\; n=9\\
U-(1/2)J & \text{for} &n=7\; \text{and}\; n=8.
\end{array} \right.
\label{eq:Ueff}
\end{equation}
Mn has $n=5$ and thus the highest $U_\eff$ which further corroborates our conclusion of Mn resembling an atomic spin $S=5/2$ with nearly frozen valence and is in line with the discussion of Kanamori type Coulomb interactions in Ref. \cite{Werner_2009,de_medici_PRB2011}. In contrast to the Kanamori model, here $U_\eff$ also varies between the non-half filled cases $(n=6,...,9)$. Most importantly, we find that the $d^6$ and $d^9$ atomic configurations yield the smallest $U_\eff$ and are most susceptible to valence fluctuations. This gives a hint towards mixed valence behavior of the Fe and Ni adatoms, which is substantiated by our QMC results.

For the Fe $d_z^2$ orbital, there are no well defined upper Hubbard bands but only spectral peaks above $E_F$ which extend to or even below the $E_F$ (Fig.\ref{fig:orbdos} b)  and d)). We further find that this overall structure of the spectra remains stable also at larger fillings, like $n=6.4$ shown in Fig.\ref{fig:orbdos}(d). In this entire range ($6<n<6.4$) of occupancies \footnote{We note that $n=6.4$ corresponds to the occupancies naively extracted from spin-polarized DFT-GGA as well as GGA+U calculations with with $U=2$\,eV up to $5$\,eV, whereas for $n=6$ best quantitative agreement of experimental and calculated spectra is achieved.}, the Fe adatoms are in a mixed valence situation.

The experimental spectrum of Ni consists mainly of a broad peak below the Fermi level and we find good agreement between the QMC calculations and the measured $3d$ spectra of Ni adatoms for occupancies $8.3 \gtrsim n \gtrsim 8.7$. \footnote{Towards integer occupancies like $d^8$ or $d^9$ a distinct satellite peak forms at $-2.5$eV resembling the feature present in the Hubbard I spectra shown in Fig.\ref{fig:spectra}(d). Since the experimental data show a minimum around this energy the filling of the Ni $3d$ shell has to be in between $n=8$ and $n=9$.} In this range, the Ni spectra obtained from our QMC simulations and the experiments are qualitatively more similar to the "non-interacting" GGA density of states than to the ED spectra shown in Fig.\ref{fig:spectra} d). There are no well defined upper Hubbard bands in any of the Ni orbitals but only broad spectral weight distributions above $E_F$ which extend below $E_F$. This points towards a mixed valence situation for also for Ni.

In this respect Fe and Ni are very different from Co: For Co, our experiments (feature 3) and calculations show a quasi particle peak  at the Fermi level, which is well separated from clearly formed upper and lower Hubbard bands. There are, thus, less charge fluctuations for Co on Ag (100) and so this system comes closer to the (multiorbital) Kondo limit. Therefore, our results confirm the interpretation of low energy resonances in STM spectroscopy experiments of Co on Ag(100) in terms of a Kondo effect \cite{wahl_2004}. The QMC results further show that all Co orbitals are involved in the quasiparticle resonance (Fig. \ref{fig:orbdos} a,c). Thus, excitations of the orbital degree of freedom must be available at energies on the order of our simulation temperature $1/\beta=0.05$\,eV. This is similar to the case of Co on Cu (111) \cite{Surer_2012}. 

For Co we find best agreement of calculated and measured spectra in the range $n=7.8-7.9$ (Fig.\ref{fig:orbdos} a,c). 
Thus, the Co is closer to a $d^8$ than to a $d^7$ configuration which supports recent coupled cluster calculations \cite{carter_2009}. We note that a prediction of the Co valency based on DFT type approaches can be misleading: LDA+U calculations with $2$\,eV$<U<5$\,eV yield an occupancy of the $3d$ shell between $n=7.0$ and $7.2$.

The spectral function of Mn (a group VII-element) could be well understood assuming a filling of $n\approx 5$ for the Mn $3d$ orbitals. However, noble metals like Cu (group XI-elements having 4 electrons per atom more than the corresponding group VII elements) have an almost full $d$ shell (i.e. $n\approx 10$) due to one electron from the $4s$ orbitals being promoted to the $3d$ orbitals. If this promotion of one electron from the $4s$ to the $3d$ orbitals would occur homogeneously, the $3d$ occupancy should increase by 1.25 electrons between each two atoms of the $3d$ series under investigation. Such an increase is in line with our results and the mixed valence behavior for Fe ($6<n<6.4$) and Ni ($8.3 \gtrsim n \gtrsim 8.7 $) but not with Co which comes closer to the Kondo limit of nearly frozen $d^8$ valence. 

\textit{In summary}, our joint experimental and theoretical study shows that Hund's exchange controls the physics of $3d$ adatoms on the surfaces of Ag (100). It fosters the formation of multiplets, determines multiplet splittings and modulates effective charging energies. Our results show that any realistic description of magnetic nanosystems should account for these manifestations of Hund's exchange. Particularly the mixed valence behavior of Fe and Ni challenges discussions of transition metal based nanomagnetic structures or impurity systems in terms of spin-only models. The situation is complex and challenging. Further spectroscopy studies, including photoelectron spectroscopy with higher energy resolution and as a function of temperature, will be useful to fully describe the nature of general excitation spectra of magnetic impurities on surfaces.

We thank A. Belozerov and A. Poteryaev for stimulating discussions. Support from SFB 668, LEXI Hamburg, the Slovenian Research Agency under Research Program no. P2-0377 and computer time at HLRN are acknowledged. 

\bibliography{thebib}

\end{document}